\documentclass
[aps,pre,epsfig,twocolumn,floats,superscriptaddress
,amssymb ,amsmath ,color,showpacs] {revtex4}
\usepackage{graphicx}
\usepackage{bm}
\usepackage{color}
\begin{document}
\newcommand{\kt}{k_{\rm B}T}
 \newcommand{\omegas}{\omega_{\rm s}}
\newcommand{\Omegas}{\Omega_{\rm s}}
\newcommand{\etas}{\eta_{\rm s}}
\newcommand{\thetas}{\theta_{\rm s}}
\newcommand{\xs}{x_{\rm s}}
\newcommand{\bfq}{{\bf q}}
\newcommand{\R}{{\cal R}}
\title{Effective interactions between inclusions in complex
fluids driven out of equilibrium
}
\author{Denis Bartolo}
\author{Armand Ajdari}
\affiliation{
Laboratoire de Physico-Chimie Th\'eorique,
UMR CNRS 7083,
ESPCI, 10 rue Vauquelin,
F-75231 Paris Cedex 05, France.}
\author{Jean-Baptiste Fournier}
\affiliation{
Laboratoire de Physico-Chimie Th\'eorique,
UMR CNRS 7083,
ESPCI, 10 rue Vauquelin,
F-75231 Paris Cedex 05, France.}
\affiliation{F\'ed\'eration de recherche FR CNRS 2438 ``Mati\`ere
et Syst\`emes complexes'',
F-75231 Paris Cedex 05, France.}
\begin{abstract}
The concept of fluctuation-induced effective
interactions is extended to systems driven out of equilibrium. We
compute the  forces experienced by macroscopic objects immersed in
a soft material driven by external shaking sources.
We show that, in contrast with equilibrium Casimir forces induced by thermal
fluctuations, their sign, range and amplitude depends on
specifics of the shaking  and can thus be tuned. We also comment
upon the dispersion of these shaking-induced forces, and discuss
their potential application to phase ordering in soft-materials.
\end{abstract}
\maketitle
\section{INTRODUCTION}
A prominent issue
in soft condensed matter physics is the understanding of the equilibrium
phase behavior of mesoscopic particles immersed in complex
fluids: colloidal suspensions~\cite{colloid}, liquid droplets in
liquid crystals~\cite{inclusionsnem}, inclusions in lipid
membranes~\cite{prot}, charged particles in electrolytes
~\cite{electrolytes},... A standard and fruitful procedure to
describe the large scale properties of inclusions in a soft
medium consists in integrating out the numerous "solvent" degrees
of freedom via ensemble averaging~\cite{colloid,onsager}. The
interaction between the embedded particles is then described by
``effective potentials". The latter can modify the genuine
interparticles interactions (Coulomb, van der Waals,...) or give
rise to entirely new effects. For instance, in some (equilibrium)
cases, the external objects do not modify the ground state energy
of the medium but only alter its thermal-fluctuations spectrum.
The resulting entropic effective interaction has consequently an amplitude
proportional to the thermal energy $\kt$ and a range comparable
to that of the correlations of the fluctuations of the medium
\cite{raminrmp,aa}. Such fluctuation-induced (i.e. entropic)
interactions are commonly referred to as thermal Casimir
interactions in analogy with their famous quantum
equivalent~\cite{casimir}.

More recently, many experimental studies have reported the
organization of particles embedded in fluids when the latter are driven out of
equilibrium by the application of external
fields:~\cite{vibrated,spinner,pompes,grier}.
Extensions of the concepts of "effective potential" and "entropic forces"  to out of equilibrium
situations have been scarce~\cite{grier,toddbrenner,hastings}.
Indeed effective potentials can not be simply
derived from a free energy in an out-of-equilibrium context,
and only the instantaneous force acting on the external objects
for a given configuration of the medium can be
defined. The
effective interactions between the host objects should then strongly depend on the
dynamics ruling the temporal evolution of the medium
(in contrast to the equilibrium Casimir case).
\begin{figure}
\includegraphics[height=5cm]{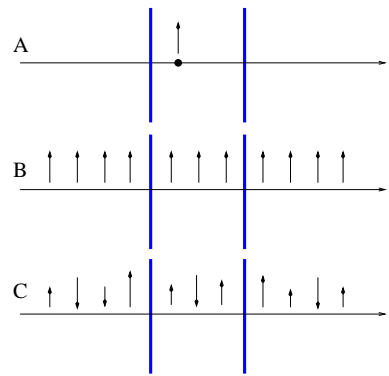} \caption{Sketch of the
three shaking processes discussed in the paper. (A): A point-like monochromatic
shaking source. (B): Uniform monochromatic shaking. (C):
Spatially uncorrelated colored noise.}\label{shakers}
\end{figure}

In this paper, we attempt to extend the paradigm of Casimir
effective interactions to objects immersed in soft systems
"shaken" by external energy sources that can not be a priori
modeled by heat baths. We
use a model, detailed in section II, where both the medium and the objects are very simple.
(i) The hosting medium is described by a scalar free field $\phi$ living in a $d$
dimensional space ($d=1,2,3$): for instance, $\phi$ could
represent the height profile of a fluid interface ($d=2$) or a
contact line ($d=1$)~\cite{interface}, or the
angular deviation of the director of a nematic liquid crystal
($d=3$)~\cite{prostdegennes}. (ii) The external objects are
two identical rigid parallel plates that enforce a zero field on their
surfaces (Dirichlet boundary conditions).
Among the numerous possible choices for the  shaking sources, we
explore three different cases of experimental relevance
(c.f. Fig.~\ref{shakers}):  (A) a localized monochromatic shaking,
(B) an uniform monochromatic shaking, (C) a spatially uncorrelated
colored noise. All these shaking induce effective
interactions between the plates, which we characterize in section III by
 the corresponding average forces, discussing their sign, amplitude and range. We compare
these features with those of the usual thermal Casimir effect.
The detail of the calculations and a more thorough analysis
of the shaking-induced interactions are gathered in section IV.
This more technical section also includes some comments
on the time dependence and fluctuations of these shaking-induced forces.
Section V ends the paper with a synthetic summary of our main results
and an outlook on possible applications.
\section{Model and notations.}
\subsection{Energetics and Force on the plates :}
\begin{figure}
\includegraphics[height=4cm]{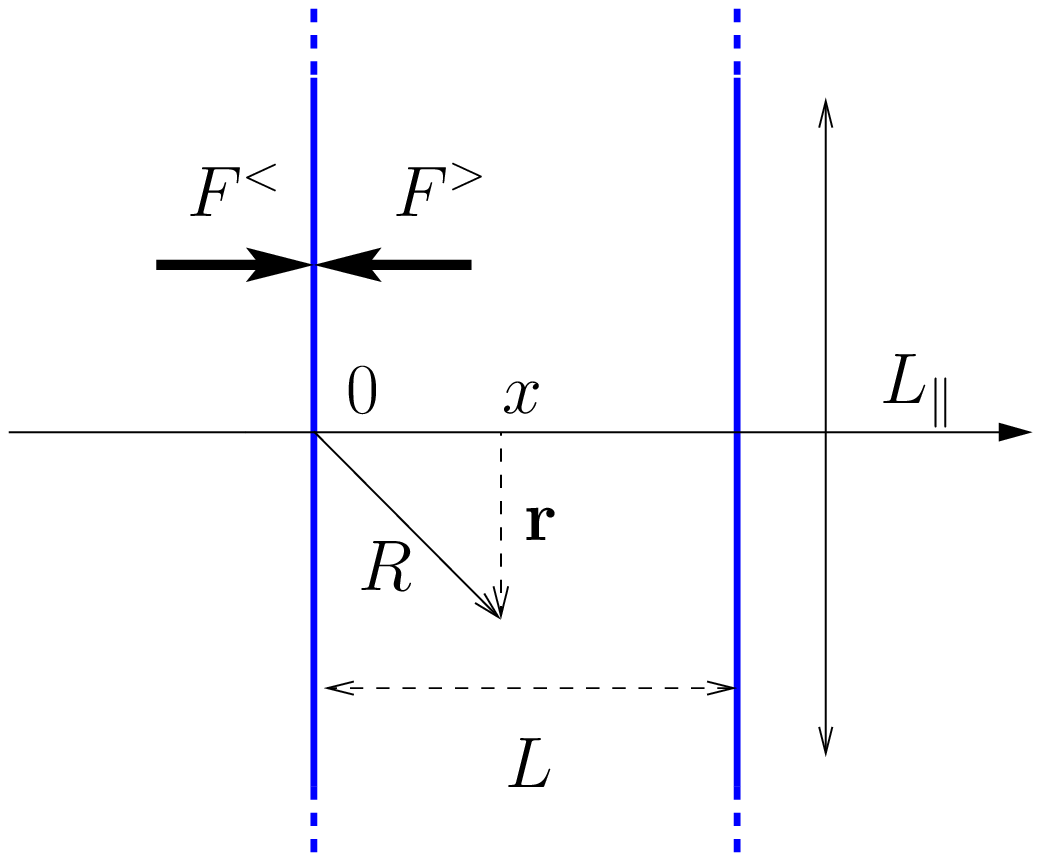}
\caption{Two parallel
plates, perpendicular to the $x$ axis, separated
by a distance $L$ shorter than their lateral extension
$L_{\parallel}$.}\label{geometry}
\end{figure}

We consider two plates in a soft medium separated by a distance
$L$ much shorter than their lateral extension $L_{\parallel}$ as
depicted in Fig.~{\ref{geometry}}. Their
thickness plays no role in all that follows and will be
set to $0$. The soft medium is modeled by a scalar field
$\phi$ associated to the elastic Hamiltonian~:
\begin{equation}
{\cal H}=\frac{\kappa}{2}\!\int\!\!d^d{\vec{R}}\,\left
[\nabla \phi(\vec{R})\right]^{2},
\label{H}
\end{equation}
with $\kappa$ the elastic modulus and $\vec{R}=(x,{\bf
r})$ where $x$ is the coordinate normal to the
plate, see Fig.~\ref{geometry}. $\phi$ is taken dimensionless.

We restrict ourselves to strong plates-field interactions,
modeled by Dirichlet boundary conditions
(DBC) on the plates: $\phi(x=0,{\bf r})=\phi(x=L,{\bf r})=0$.

We first focus our attention on the force $F^>$ that the soft
medium exerts on the right hand side of the plate located at
$x=0$. For a given configuration of the elastic field, $F^>$  is
given by the integral over $\bf r$  of the stress tensor
component $T_{xx}({\bf r})$ ~\cite{landau}. The components
$T_{ij}$ of this stress tensor can be obtained by
the virtual displacement method taking into account the
DBC~\cite{aa}. In the present geometry, $T_{xx}({\bf r})=-\frac{1}{2}\kappa\left
[\partial_x\phi(0^+,{\bf r})\right]^2$, so that
\begin{equation}
F^>=-\frac{\kappa}{2}\int\!d{\bf r}\,[\partial_x\phi(0^+,{\bf
r})]^2.\label{force}
\end{equation}
Note that this force pushes the plate toward the negative
$x$ direction whatever the field conformation.

The net force $F$ on the plate is the algebraic sum of the contributions of the
two sides:
$F\equiv F^>+F^<$. The force $F^<$ is given by
a formula similar to Eq.~(\ref{force}), so that:
\begin{equation}
F=- \frac{\kappa}{2}\int\!d{\bf r}\,[\partial_x\phi(0^+,{\bf
r})]^2  + \frac{\kappa}{2}\int\!d{\bf r}\,[\partial_x\phi(0^-,{\bf
r})]^2.
\label{forcetot}
\end{equation}

\subsection{Dynamics :}
For sake of clarity all  calculations are
performed within the simplest local and strongly dissipative
dynamical scheme: the $\phi$
field evolves under the application of an external "shaking"
source $\eta$ as~:
\begin{eqnarray} \gamma
\partial_t\phi&=&-\frac{\delta {\cal H}}{\delta\phi}
+\eta,\nonumber\\&=&\kappa\nabla^2\phi
+\eta,\label{dynamique}\\
\phi(0,{\bf r},t)&=&\phi(L,{\bf r},t)
=0
\label{cl}.
\end{eqnarray}
Here $\gamma$ is a generalized friction coefficient.

A description of the
dynamics of the plates is beyond the scope of this paper, and we assume
that they are fixed. $L$ is thus a constant. After partial
Fourier transform with respect to $t$ and $\bf
r$, equations (\ref{dynamique},\ref{cl}) can be recast
into:
\begin{equation}
\phi_{\omega,\bf
q}(x)=\int_0^L\!\!dx'\,{\cal R}_{\omega,\bf
q}^>(x,x')\eta_{\omega,\bf q}(x'),
\label{reponse}
\end{equation}
where the
response function ${\cal R}^>$ corresponds to the diffusion
kernel in the slab geometry with DBC and with a diffusion constant $\kappa/\gamma$. 
The Fourier transforms used in this paper are defined by
$f_{\bf q}\equiv\int\!d{\bf r} \,f({\bf r})\exp(i{\bf q.x})$ and
$g_{\omega}\equiv\int\!dt \,g(t)\exp(i\omega t)$.

A generalization of our results obtained with (\ref{dynamique},\ref{cl})
to other slow dynamics that obey an equation of the
form~(\ref{reponse})  with a different kernel will be commented
upon throughout the text. This generalization is simple provided
the dynamic linear response of $\phi$
relates spatial and temporal scales
through an algebraic relation $l_\omega\sim\omega^{-1/z}$, where
$l_\omega$ is the spatial extension of the elastic
distortion resulting from a localized periodic shaking of
pulsation $\omega$, and $z$ is the so-called dynamical exponent.
The diffusive model described above by Eq.  (\ref{dynamique},\ref{cl})
indeed fits in this picture with
$z=2$ and:
\begin{equation}
l_\omega=\left[(\gamma/\kappa)\omega\right]^{-1/2}
\label{lomega}
\end{equation}
\section{Average Forces induced by three kinds of shaking}
To set a reference for further comparison,
we first recall the expression of the average thermal force
$F_{\rm Casimir}$ on the plate at $x=0$ if the whole medium
is thermally excited by a heat bath imposing a temperature
$T$~\cite{kardarlee}.
The simplest derivation consists in
computing the total free energy of the system for a given
interplate distance $L$, and taking its derivative with respect
to $L$. An alternative approach is to consider
the model Langevin dynamics for the field given by
Eq.(\ref{dynamique},\ref{cl}), with $\eta$ a thermal Gaussian
white noise of zero mean and of variance
$\langle\eta(\vec{R},t)\eta(\vec{R}',t')\rangle=2\gamma\kt\delta(\vec{R}
-\vec{R}')\delta(t-t') $.
Averaging over realizations of the noise
leads to the same universal
expression for the net thermal Casimir
force~\cite{aa}:
\begin{equation} F_{\rm Casimir}=A_{d}\,\kt
\frac{L_{\parallel}^{d-1}}{L^{ d } },
\label{casimirforce}
\end{equation}
with $A_{d}=(d-1)\Gamma(d/2)\zeta(d)/(4\pi)^{d/2}$.
This universal mean force describes a long range attraction
that depends on the temperature of the heat bath
but not on specific material properties such as $\kappa$ and
$\gamma$.

We now calculate the average forces generated by the three
non-thermal shaking processes mentioned in the introduction, with a field $\phi$
that evolves according to Eq.~({\ref{dynamique},\ref{cl}}).

\subsection{Average force
induced by a localized monochromatic shaking}
We start with the simplest choice for the shaking
in Eq.~(\ref{dynamique}), namely a point-like monochromatic
shaker oscillating at a pulsation $ \omega_{\rm
s}$, located at an equal distance $L/2$ from the two
plates:
\begin{equation}
\eta(\vec{R},t)=\eta_{\rm s}a^d\cos(\omega_{\rm
s}t) \delta(x-L/2)\delta(\bf r),
\label{etaa}
\end{equation}
The noise amplitude has been arbitrarily split into the product
of a microscopic volume $a^d$ where the medium is
shaken, times a shaking amplitude $\etas$
that has the dimension of an energy
density.

Such a process could be, for example, the result of
the forcing of a microscopic "active" particle located between
the two plates (e.g. a magnetic particle under the
influence of an external oscillating magnetic field),
in which case $a$ is typically fixed whereas $\etas$ and $\omega_{\rm s}$
can be externally tuned.

\begin{figure}
\includegraphics[width=8cm]{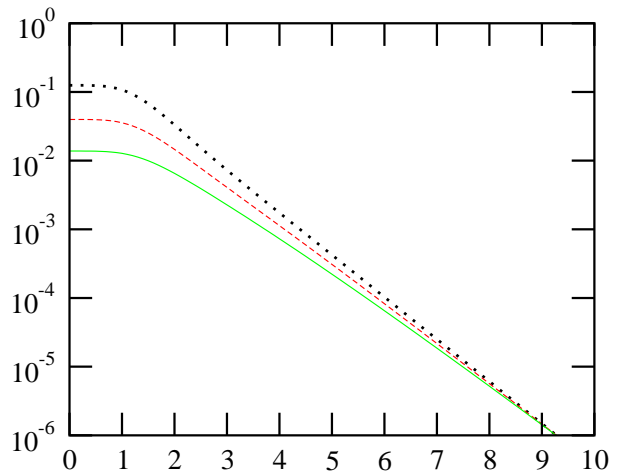}
\caption{{\it Localized monochromatic shaking}. Log-linear plot of
the scaling function $\Upsilon_{\rm A}(1/2,
L/l_{\omegas } )$ against $L/l_{\omegas}=(\gamma/\kappa)^{1/2}\omegas^{1/2}L$,
for various space dimensions:
$d=3$ full line, $d=2 $ dashed line, and $d=1$ dotted
line.}\label{upsilonaplot}
\end{figure}

The net force $F$ felt by the plate at $x=0$,
reduces to its right hand side contribution $F^>$ since the outer
medium remains undistorted ($\phi(x<0)=0$). Moreover
the geometry is here totally symmetrical so that the force felt
by the plate at $x=L$ is: $-F$. Denoting $F_{\rm A}^>$
the time averaged force for this process, and anticipating on
the detailed discussion of section IV-A, we use
Eq.~(\ref{upsilona}) to write:
\begin{equation}
F_{\rm A}^>=-\frac{(\etas a^d)^2}{\kappa}
\frac{1}{L^{d-1}}\Upsilon_{A}\left(\frac{1}{2},
\frac{L}{l_{\omegas } } \right ).
\label{FA}
\end{equation}
Again
$l_{\omegas}=[(\gamma/\kappa)\omegas]^{-\frac{1}{2}}$ is the
dynamical length associated with the shaking pulsation $\omegas$.
The adimensional function $\Upsilon_{\rm A}(1/2,L/l_{\omegas})$
is plotted in Fig.~\ref{upsilonaplot}, and its general
expression (including the shaker position dependence) is given
below in  Eq.~(\ref{upsilona}).

Two different limit regimes can
be identified: If $L/l_{\omegas} \ll1$, the shaking period is much
larger than the relaxation of any interplate field excitation, so
that the elastic deformations generated at $x=L/2$  propagate
along the $x$ axis up to the plates surfaces. In this quasistatic
limit, the time averaged force $F_{\rm A}^{>}$ felt by the plate
at $x=0$ is comparable to the elastic force on the plates under
the application of a constant static ``effective'' perturbation
$\frac{1}{\sqrt{2}}\etas a^d\delta(x-L/2)\delta({\bf r})$.
The latter induces static deformations of lateral extension $L$
on the plate. Thus, $F_{\rm A}^>$ scales as
$\etas^2a^{2d}/(\kappa L^{d-1})$ independently of the shaking
period. In contrast, the dynamic length is much
smaller than the plate-shaker distance if $L/l_{\omegas} \gg1$. The
elastic deformations are exponentially screened before reaching
the plates surfaces. Therefore, the force amplitude decays
exponentially as shown in Fig.~\ref{upsilonaplot}.

In summary, a localized periodic shaking  induces an
effective interplate interaction.
Whereas $F_{\rm
Casimir}$ is universal and leads to an attraction between the
two plates, the interaction obtained here is repulsive and does
depend both on the elastic modulus of the surrounding``fluid" and
on its dynamics (through $l_{\omegas}$). Finally, both the
range and the amplitude of the present shaking-induced
force can be externally tuned by varying the pulsation $\omegas$
and the amplitude $\etas$ of the shaking. The dependence of the force on the shaker position as
well as its fluctuations will be discussed in section IV.
\subsection{Average force induced by a uniform monochromatic
shaking}
We now suppose that the elastic medium is everywhere periodically
and homogeneously shaken:
\begin{equation}
\eta(\vec{R},t)=\etas\cos(\omegas t).
\label{etab}
\end{equation}
Posible experimental realizations include: the homogeneous and periodic shaking of a metallic elastic line using a magnetic field~\cite{couder} or the driving of an interface between two dielectic liquids using AC electric fields~\cite{xx}. One may also use  flexoelectric effects in liquid crystal to generate such a shaking process~\cite{prostdegennes2}.
\begin{figure}
\includegraphics[width=8cm]{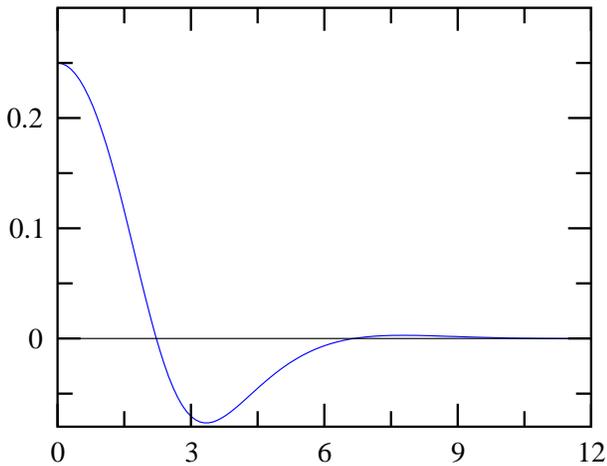}
\caption{{\it Uniform monochromatic shaking}. Linear plot of the scaling
function $\Upsilon_{\rm B}(L/l_{\omegas })$ as a function
of $L/l_{\omegas }=(\gamma/\kappa)^{1/2}\omegas^{1/2}L$
.} \label{upsilonbplot}
\end{figure}

By symmetry the two plates
feel opposite instantaneous forces.
The exact expression for the time averaged force
$F_{\rm B}$ on the $x=0$ plate  is computed in section IV:
\begin{equation} F_{\rm B}=\frac{\etas^2
L_{\parallel}^{d-1}}{\kappa}\,l_{\omegas}^2 \Upsilon_{\rm
B}\left(\frac{L}{l_{\omegas}}\right).
\label{FB}
\end{equation}
The dimensionless quantity
$\Upsilon_{\rm B}$, see Eq.~(\ref{upsilonb}),  is plotted in
Figure~\ref{upsilonbplot}: it is non-monotonic and changes sign periodically.
Given the plates geometry
and the shaking translational invariance,
the net normal stress  $F_{\rm B}/L_{\parallel}^{d-1}$ depends neither
on the plates lateral extension $L_{\parallel}$ nor on the dimension
$d$. The fast and slow shaking asymptotics can thus be
qualitatively inferred from dimensional analysis.  If the
medium is slowly shaken ($L/l_\omega\ll1$), the net force
amplitude diverges as $l_{\omegas}^2\sim1/\omegas$.  First,
notice that a constant uniform source $\etas$ between the plates
induces an homogenous normal stress on the inner side of the
plate  ($x=0^+$) that scales necessarily as $F_{\rm
B}^>/L_{\parallel}^{d-1}\sim \etas^2\kappa^{-1}L^2$. Second, on
the left hand side of the plate ($x<0$), $l_{\omegas}$ is the only length
scale available to construct the normal stress $F_{\rm
B}^</L_{\parallel}^{d-1}\sim\etas^2\kappa^{-1}l_{\omegas}^2$. In
this slow shaking regime this latter contribution dominates,
so that $\Upsilon_{\rm B}$ plateaus to a finite
value. Turning now to the $L/l_\omega\gg1$ limit, if the interplate distance goes to infinity,
the two sides of the plate face identical semi infinite media.
Consequently, they are pushed in opposite directions with the
same amplitude and thus feel on average no net force. So, in this
limit $L/l_\omega\gg1$  that also corresponds to fast shaking at fixed $L$,
$F_{\rm B}$ decays to zero,
as observed in Fig. (\ref{upsilonbplot}).

Returning to the most remarkable features,
Fig.~\ref{upsilonbplot} and Eq.~(\ref{FB}) show that the
fluctuations of a uniformally shaken medium induce effective
interplate forces that strongly differ from $F_{\rm Casimir}$.
They are indeed completely tunable unlike their equilibrium
analog. Their sign, amplitude and range can all be controlled
externally by varying the shaking period and amplitude.
\subsection{Average force  induced by a stochastic shaking: colored noise}
Finally the
case of stochastic shaking sources is considered.
As mentioned above, if $\eta$ is the white Langevin noise
modeling the coupling to a heat bath, the resulting field
fluctuations induce thermal Casimir interactions between the plates.
Any other noisy shaking drives the system out of
equilibrium. Numerous examples of elastic materials shaken
by ``active" stochastic processes are provided by biological
systems, e.g. active membranes~\cite{pompes}, actin-myosin
gels~\cite{gels}, etc.

\begin{figure}
\includegraphics[width=8cm]{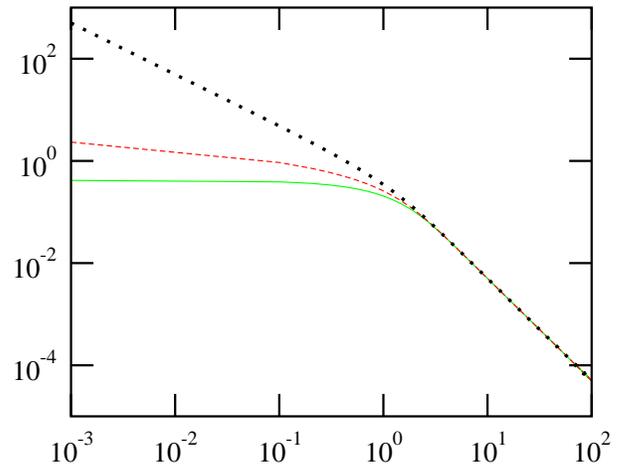}
\caption{{\it Colored noise}. Log-log plot of the ratio $\Upsilon_{\rm C}(L/l_{\Omegas})/A_{d}$ versus $ L/l_{\Omegas}=(\gamma/\kappa)^{1/2}\Omegas^{1/2}L $. The full
line corresponds to $d=3$, the dashed line to $d=2$
and the dotted line to $d=1$.}
\label{upsiloncplot}
\end{figure}

Without attempting to describe accurately the specifics of the noise in a
given system, we focus here on the simple case of a
spatially uncorrelated colored
noise of zero mean, seeking for the influence of its color on the
interplate force generated.
We describe the
fluctuations of this noise by:
\begin{eqnarray}
&& \langle\eta(\vec{R},t)\eta(\vec{R}',t')\rangle=
\etas^2 a^dS(t-t')\delta(\vec{R}-\vec{R}'),
\label{etac}
\end{eqnarray}
where the correlation $S(t)$ is defined by its Fourier transform
$S_{\omega}$, also referred to as the noise power spectrum:~
\begin{equation}
S_{\omega}=\frac{\Omegas}{\Omegas^2+\omega^2}.
\label{powerspec}
\end{equation}
In Eq.~(\ref{etac}) $a$ is a microscopic distance chosen so that
the mean squared noise  satisfies $\langle
\eta^2(\vec{R},t)\rangle=\etas^2$. Again $\etas$  has the
dimension of an energy density.
The white noise limit corresponds to $\Omegas\to \infty$, with
$\etas^2 a^d\Omegas^{-1}$ kept constant.

The mean net force $F_{\rm C}$ on the $x=0$ plate is
exactly computed in section IV-C:
\begin{equation}
F_{\rm
C}=\frac{\eta_{\rm s}^{2}a^d}{\kappa}\frac{L_{\parallel}^{d-1}}{
L^{d-2}}\Upsilon_{\rm C}\left(\frac{L}{l_{\Omegas}}\right).
\label{FC}
\end{equation}
The dimensionless ratio $\Upsilon_{\rm C}/A_{d}$ is plotted in
Fig.~\ref{upsiloncplot}, using Eq.~(\ref{upsilonc}). We recall that
$A_d$ measures the amplitude of the thermal Casimir force~ ($A_{d}=(d-1)\Gamma(d/2)\zeta(d)/(4\pi)^{d/2}$).

Postponing a more involved discussion to next section, we
emphasize here three points. First,
the color of the noise has no influence on the sign of the force: the two
plates attract each other whatever $\Omegas$ and $\etas$.
Second, $\Upsilon_{\rm C}$ decreases monotonically with
$L/l_{\Omegas}$, in other words the larger the noise
correlation time $\Omegas^{-1}$, the stronger the force felt by
the two plates.
Third, Fig. (\ref{upsiloncplot}) indicates that $\Upsilon_{\rm
C}(u)/A_d$ decays as $1/(2u^2)$ when
$u\gg1$. This implies that the attraction is long-ranged and decays only
algebraically with the interplate distance.
A connection with the Casimir result appears in this regime:
if one sets
$\etas^2a^d\Omegas^{-1}=2\gamma\kt$,
then in the limit white noise limit $\Omegas \rightarrow \infty$, $F_{\rm C}$ converges as
expected to $F_{\rm Casimir}=A_{d}\kt L_{\parallel}^{d-1}/L^d$.

To conclude, we have shown that Casimir-like attractions
can be induced by colored noisy shaking. These attractions decay
algebraically at long distance as in the thermal
case and contrary to the two former monochromatic cases.
Their amplitude depend
explicitly on the material properties and their scaling on the
noise color.
\section{Shaking-Induced Forces : Detailed analysis.}
In this section we provide a more detailed analysis of
the interactions induced by the three kind of shakings:
in addition to the explicit derivation
of the average of the corresponding forces, we
also comment on their temporal fluctuations.

We start with the derivation of the tool that will allow us to compute
the forces on the $x=0$ plate in all situations, namely the linear
response function $\partial _x {\cal R}$ of the field's gradient at the
surface of the plate to a point-like shaker in the medium (by
definition, the kernel $\cal R$ is the solution of the dynamic equations
(\ref{dynamique},\ref{cl}), with
$\eta(\vec{R},t)=\delta(\vec{R}-\vec{R}')\delta(t-t')$). We compute this
quantity using an image method which conveys an intuitive
picture~\cite{MF}: as in electrostatics, these equations can be solved
replacing the constraint at the boundaries (\ref{cl}) by the addition of
an ad hoc distribution of image sources outside the integration domain.

\begin{figure}
\includegraphics[width=8cm]{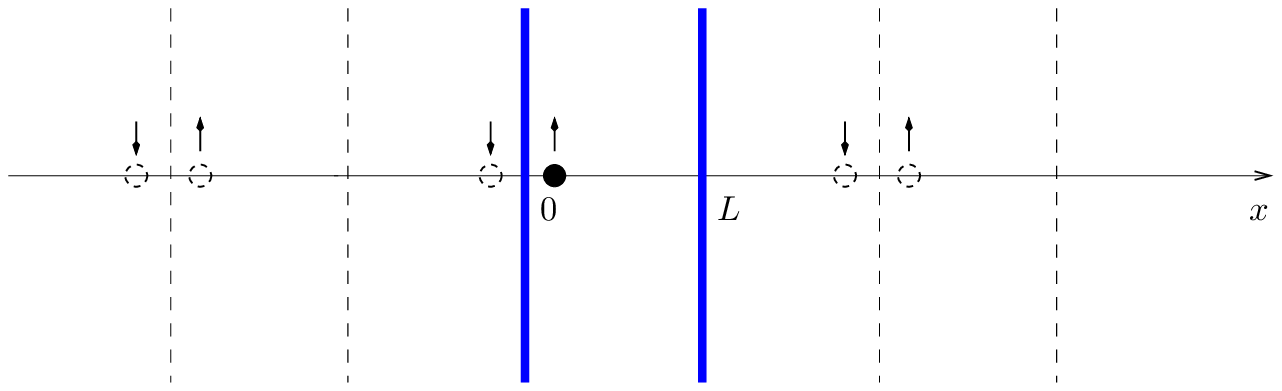}
\caption{Distribution of the image sources for the
two plates geometry and Dirichlet boundary
conditions. The filled circle is the original source
and the dashed circles its mirror images}\label{images}
\end{figure}

We consider first a shaking source located
in the interplate region $0<x'<L$ and compute the response
$\partial _x {\cal R}^> $ on the $x=0^+$ side of the plate.
In the present slab geometry, the images are the mirror images of the
original source through virtual reflecting planes
located at $x=nL$ as depicted in Fig.~\ref{images}.
Consequently, the response
$\R_{\omega,\bfq}^>(x,x')$ can be expressed as an infinite sum
over all the images contributions:
\begin{eqnarray}
\R_{\omega,\bf q}^>(x,x')&=& \sum_{n\in\mathbb{Z}}\R_{\omega,\bf
q} \left(2nL+x'-x\right)\label{imagesum}\\
&-&\sum_{n\in\mathbb{Z}}\R_{\omega
, \bf q}\left(2nL-x'-x\right),\nonumber
\end{eqnarray}
where $\R$ the diffusion kernel in an infinite domain :
\begin{eqnarray}
&&\R_{\omega,\bf
q}(x,x')=\frac{\exp\left({- \sqrt{q^2+i\,l_\omega^{-2}}\left
|x-x'\right|}\right)}{2\kappa\sqrt{q^2+ i\,l_\omega^{-2}}
},
\end{eqnarray}
with $\sqrt{i}\equiv-1$. Performing the  sum~(\ref{imagesum}) and
taking the derivative with respect to $x$ yields:
\begin{equation}
\partial_x\R_{\omega,\bfq}^>(0,x')=
\frac{\sinh\left [ \sqrt{(qL)^2+i(L/l_{\omega})^2}( 1 -
x'/L)\right]}{\kappa \sinh \left [
\sqrt{(qL)^2+i(L/l_{\omega})^2}\right ]}\\.\label{gradR}
\end{equation}

A simpler calculation can be carried out if the source
is in the negative
$x$ region ($x'<0$). In this case the only image source is
the symmetric in the $x=0$ plane of the original source,
so that the response
of the field gradient on the $x=0^-$ side of the plate is described by:
\begin{equation}
\partial_x\R_{\omega,\bfq}^<(0,x')=\frac{e^{-
\sqrt{(qx')^2+i(x'/l_{\omega})^2}}}{\kappa}.
\label{gradRgauche}
\end{equation}
\subsection{Localized Periodic Shaking.}
If the only driving is a single monochromatic shaker
 located between the plates at $x=\xs$,
\begin{equation}
\eta=\etas a^d\delta(x-\xs)\delta({\bf r})\cos(\omegas t),
\label{etaa2}
\end{equation}
then the general
expression of the net force acting on the $x=0$
plate is simply obtained from the definition of the response
function $\partial_{x}{\cal R}^>$:
\begin{eqnarray}
F(t)=\frac{-\kappa(\eta_{\rm s}a^d)^2}{4}\left[  \int_{\bf q}
\! \left |\partial_x {\R}_{\omegas, {\bf q}}^>(0,\xs)\right|^2
\right ]\times\label{FAt}\\ \left [1+\cos(2\,\omegas
t)\right],\nonumber \end{eqnarray}
with the short hand notation
$\int_{\bf q}\equiv \int d{\bf q}/(2\pi)^{d-1}$
\cite{phase}.

There is obviously no force on the left side of the plate,
so that using (\ref{gradR}) we obtain the exact expression
for the instantaneous force on this plate.
Its time average $F_{\rm A}^>$ describes a repulsion
between the plates:
\begin{equation}
F_{\rm
A}^>(\xs,\omegas)=-\frac{(\etas a^d)^2}{\kappa L^{d-1}}
\Upsilon_{\rm A}\left (\frac{\xs}{L},\frac{L}{l_{\omegas}}\right)
\label{FAxs}
\end{equation}
with,
\begin{equation}
\Upsilon_{\rm A}(u,v)=\frac{1}{4}\int_{\bfq}\left
|\frac{\sinh\left [\sqrt{q^2+iv^2}\,(1-u)\right] } {
\sinh(\sqrt{q^2+iv^2})} \right |^2.\label{upsilona}
\end{equation}
We now give a closer look at the asymptotics corresponding
to the fast and slow shaking limits.\\

\begin{figure}
\includegraphics[width=8cm]{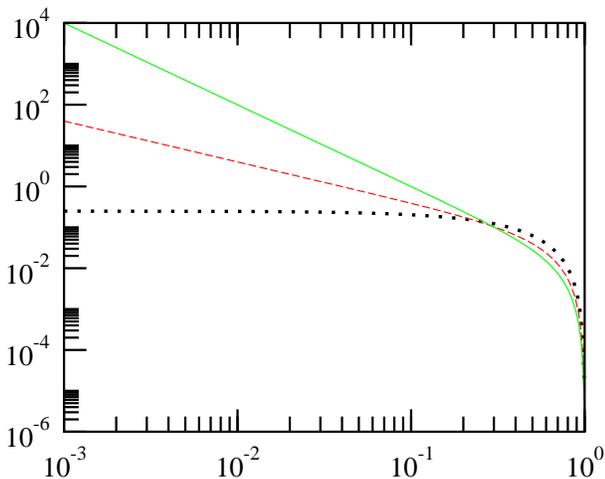}
\caption{Log-log plot of the scaling function $\Upsilon_{\rm A}(\xs/L,0)$ versus
$\xs/L$.
This corresponds to case A in the quasi-static limit $L/l_{\omegas}\ll1$.
The full line corresponds to $d=3$, the dashed one
to $d=2$ and the dotted line to $d=1$}\label{electro}
\end{figure}
{\it Quasi-static shaking:}
We first consider the regime where the
driving period is much larger than the relaxation time of any
interplate field deformation, i.e. the dynamical length is much
larger than the interplate distance
($L/l_{\omegas}\ll1$).

The limiting value
$\Upsilon_{\rm A}(u,0)$ can be
exactly computed by integrating over the in plane ${\bf q}$ modes
before summing up all the contributions from all images in
Eqs.~(\ref{imagesum},\ref{FAt}). If $d=3$ $\Upsilon_{\rm A}(u,0)=
\partial_{u}\left[\pi\left(1-u\right)\cot\left(\pi
u\right)-2\Gamma u+ 2\left(1-u\right)\psi\left(u\right)
\right]/(32\pi)$,~with $\psi$  the Euler $\psi$-function, and
$\Gamma$ the Euler constant~\cite{gradstein}, if $d=2$,
$\Upsilon_{\rm A}(u,0)=\left[1+\pi\left( 1 - u \right) \,\cot
(\pi u)\right ]/(8\pi)$, and if $d=1$, $\Upsilon_{\rm
A}(u,0)=(1-u)^2 /4$.
The scaling function $\Upsilon_{\rm A}(\xs/L,0)$  is
plotted in Figure~\ref{electro} (log-log plot). The fast
decay for values of $\xs/L$  larger
than $\frac{1}{2}$ witnesses that the elastic distortions are
essentially occurring in the $L/2<x<L$ region. Conversely
a shaker close to $x=0$ induces strong variations of the
elastic field in the vicinity of the plate, leading
to an algebraic divergence of the force when $\xs/L \rightarrow 0$
(if $d>1$).\\
The force dependence on the shaker position, can also be
qualitatively understood thanks to an electrostatic analogy.
In the present quasi-static limit, the kernel $\R^>_{\omegas}$
corresponds to the inverse of the Laplace operator (with DBC on
the plates). Consequently $\phi$ plays the role of an
electrostatic potential, the plates mimic a grounded planar
capacitance, and the shaking source a point like particle bearing
a charge $\etas a^d$. The force on the plate is then analogous to
the electrostatic force produced by a point like
charge~\cite{jackson}. From the image expansion (\ref{imagesum})
we know that this force is identical to the one produced on a
virtual plane at $x=0$ by an infinite number of $(+\etas a^d)$
charges located at $x=2nL+\xs$ and of $(-\etas a^d)$ charges at
$x=2nL-\xs$. Hence:\\
--If $\xs<L$ (c.f
Fig.~\ref{images}): the electric field on the plate
surface is dominated by the image charges at
$|x|=\xs$, so
$F\sim-(\etas a^d)^2/(\kappa \xs^{d-1})$. Higher order
reflections produce image sources that form dipoles viewed from
$x=0$. Their subdominant contributions to the electrostatic force
actually reduce the above estimation. This implies that the
further the second plate, the bigger the  force produced by
a single shaker.\\
--if $\xs\sim L$ (c.f
Fig.~(\ref{images})): the image charges
form dipoles located at $x=(2n+1)L$. The electric field on the
plate is dominated by the influence of the dipole located at
$|x|=L$ (the closest from the $x=0$ plate). The force therefore scales
$F_{\rm A}^>\sim-(\etas a^d)^2(1-\xs/L)^2/(\kappa L^{d-1})$ .\\

{\it Fast shaking:}
\begin{figure}
\includegraphics[width=8cm]{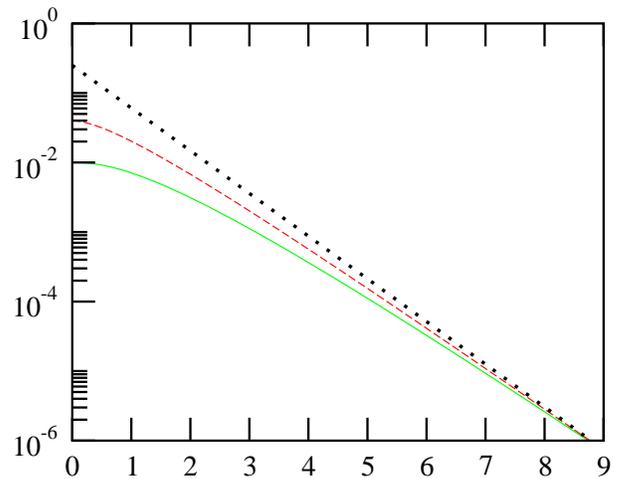}
\caption{Log-linear plot of the scaling function $\tilde{\Upsilon}_{\rm A}$.
The full line corresponds to $d=3$, the dashed line to $d=2$
and the dotted line to $d=1$}\label{fast}
\end{figure}
 If $L/l_{\omegas}\ll1$, the
electrostatic analogy only provides a partial description of the
force generation. In this case, only the  first reflexion
contribute substantially to the image
expansion (\ref{imagesum}) as if the second plated (at $x=L$) was
pushed to infinity. The only remaining length scales
are $\xs$ and $l_{\omegas}$ and the average force can be
approximated by
\begin{eqnarray} F_{\rm
A}^>\sim-\frac{(\etas a^d)^2}{\kappa
\xs^{d-1}}\tilde{\Upsilon}_{\rm A}(\xs/l_{\omegas}),
\label{FAfats}
\end{eqnarray}
with
$\tilde{\Upsilon}_{\rm A}(u)=\frac{1}{4}\int_{\bf
q}\left|\exp(-\sqrt{q^2+iu^2})\right|^2$. This
dimensionless screening factor is plotted  in figure~\ref{fast}.
If the dynamic length $l_{\omegas}$ is larger than the
distance separating the shaking source from the plate,
$\xs/l_{\omegas}\gg1$, the amplitude of the elastic field
excitations are exponentially reduced before
reaching the plate's surface.
More precisely, in this regime~:
\begin{eqnarray}
\tilde{\Upsilon}_{\rm A}(u)\sim
(u)^{\frac{d-1}{2}}\exp(-\sqrt{2}u).
\label{upsilonatilde}
\end{eqnarray}
Oppositely, $\tilde{\Upsilon}_A(\xs/l_{\omegas})$ has a non
vanishing finite limit if $\xs/l_{\omegas}$ goes to zero. $F_{\rm
A}^>$ then corresponds to the electrostatic force acting on a
conducting plate facing a single ``effective'' charge $\etas a^d$ located at
$(x=\xs,{\bf r}=0)$, so that, as in the quasi-static case
above, $F_{\rm A}^>\propto-(\etas a^d)^2/(\kappa \xs^{d-1})$.\\

{\it Fluctuations of the shaking-induced force}:
 Beyond the definition and the characterization of the average
interplate interaction, it is crucial to notice that the present
shaking induced force is a time varying quantity.
Rewriting
(\ref{FAt}) as $F(t)=F^>_{\rm A}\left[1+\cos(2\omegas t)\right]$,
it is obvious that the force experienced by the
immersed plates can not be a priori reduced to its
sole average value. As a matter of fact the fluctuations of the instantaneous force
are comparable to its average amplitude.
\subsection{Homogenous Monochromatic Shaking.}
We now consider that the soft medium is uniformally shaken by
$\eta(\vec{R},t)=\etas \cos(\omegas t)$.
$\eta$ is the  superposition of in phase shakers of
the --A-- type uniformally distributed in space. All the
elementary shakers produce periodic elastic deformations that
``interfere'' and give rise to the instantaneous forces
$F^>(t)$ (resp. $F^<(t)$) at $x=0^+$ (resp. $x=0^-$) obtained
from Eqs.~(\ref{force},\ref{reponse}):
\begin{eqnarray}
F^>(t)=-\frac{\kappa}{4}L_{\parallel}^{d-1}\left|\int_{0}^{L}
\!\!\partial_{x}{\cal R}^>_{\omegas,\bf 0}(0,x')dx'\right
|^2\times \label{FBtd}\\ \left [ 1 + \cos(2 \omegas
t+\theta^>)\right],\nonumber
\end{eqnarray}
with $\theta^>=2\arg\left[\int_{0}^{L}\!\partial_x{\cal
R}^>_{\omegas,\bf0}(0,x')dx'\right]$, and
\begin{eqnarray}
F^<(t)=\frac{\kappa}{4}L_{\parallel}^{d-1}\left|\int_{0}^{\infty}
\!\!\partial_{x}{\cal R}^>_{\omegas,\bf 0}(0,x')dx'\right
|^2\times\label{FBtg}\\ \left [ 1 + \cos(2 \omegas
t+\theta^<)\right],\nonumber
\end{eqnarray}
with $\theta^<=2\arg\left[\int_{0}^{\infty}\!\partial_x{\cal
R}^<_{\omegas,\bf0}(0,x')dx'\right]$.
At every instant these two
forces compete. They push the plate in opposite directions to
give rise to the net force $F(t)$. The latter is exactly computed
using Eqs.
(\ref{forcetot},\ref{gradR},\ref{gradRgauche},\ref{FBtd},\ref{FBtg}) ,
and after some elementary algebra~\cite{phase}:
\begin{eqnarray} F(t)&=&F_{\rm B} \left[1+\frac{\cos(2\omegas
t)}{\cos\left[L/(\sqrt{2}\,l_{\omegas})\right]}\right].
\label{FBexact}
\end{eqnarray}
The average $F_{\rm
B}$ is given by:
\begin{eqnarray} F_{\rm
B}&=&\frac{\etas^2L_{\parallel}^{d-1}}{\kappa}l_{\omegas}^2
\Upsilon _ { \rm
B}\left(\frac{L}{l_{\omegas}}\right),\label{FBbis}\\
\Upsilon_{\rm
B}(u)&=&\frac{1}{2}\frac{\cos(u/\sqrt{2})}{\cos(u/\sqrt{2})+\cosh
( u/\sqrt{2}) }.\label{upsilonb} \end{eqnarray}
The function
$\Upsilon_{\rm B}$ is plotted in section III-B, see
Fig.~{\ref{upsilonbplot}}, where its fast and slow
asymptotics are qualitatively studied.

{\it Mean force tunability:}
The tunability of the present shaking-induced force is clearly
revealed by Eqs.~(\ref{FBbis}) and (\ref{upsilonb}).
Precisely for fixed $L$: (i) The shaking pulsation $\omegas$
sets the sign of $F_{\rm B}$ to ${\rm
sign}(\cos[L/(\sqrt{2}l_{\omegas})])$, and its range to
$\sim l_{\omegas}$ (via the $\cosh$ term in (29)),(ii) then
$\etas$ allows to adjust the mean force amplitude
 arbitrarily. This second assertion is actually true
only if $L\neq\frac{\pi}{\sqrt{2}}(2n+1)l_{\omegas}$,
with $n$ integer. In such cases the plates
experience no mean net force.

{\it Fluctuations of the shaking-induced force:}
Eq.~(\ref{FBexact}) indicates actually that $F(t)$
``fluctuates'' periodically around $F_{\rm B}$ with a pulsation
$2\omegas$. Moreover the amplitude of the $2\omegas$ components
of the force is $1/\cos[L/(l_{\omegas}\sqrt{2})]$ times larger
that its mean value $F_{\rm B}$ whatever the shaking
parameters $\omegas$ and $\etas$.

{\it Shaking-induced organization:}
The two preceeding comments strongly suggest that
the homogenous shaking may be used to dynamically induce spacial
``localization'' of the plates. As a matter of fact, when
mesoscopic objects are immersed in a soft medium, their intrinsic
dynamics (ignored in the present work) necessarily filters out
 high frequency effects of external forcing. If their
response time is much larger than $1/(2\omegas)$, these ``slow''
moving objects only respond to the stationary component $F_{\rm
B}$ of the shaking-induced force.
The corresponding steady state interplate distance in the present
simple geometry is then controllable by tuning $\omegas$.
\subsection{Colored Noisy Shaking.}
The medium surrounding the
plates is now supposed to be driven by stochastic shaking
sources with zero mean and mean squared fluctuations given by
Eq.~(\ref{etac}). The noise averaged net force $F_{\rm C}$
experienced by the plate at $x=0$ is the sum of the
mean values of the two independent forces $F^>$ and $F^<$. Using
Eqs.~(12,18,20) this force is written:
 \begin{eqnarray}
 F_{\rm C}=\frac{2L_{\parallel}^{d-1}}{a^d}\int_{\omega}\!
S_{\omega}\left[ \int_{0}^{L}\!\!(F_{\rm
A}^>(x,\omega)+F_{\rm A}^<(-x,\omega))
\,dx\right. \label{FCsumFA}\\
\left .+\int_{L}^{\infty}\!\!\!F_{\rm
A}^<(-x,\omega)\,dx\right ].\nonumber
\end{eqnarray}

 {\it Exact results}: For a Lorentzian power spectrum, see
Eq.~(\ref{powerspec}), the integration over frequency and
position of all the elementary contributions $F_{\rm
A}(x,\omega)$ in (\ref{FCsumFA}) yields:
\begin{eqnarray} F_{\rm
C}&=&\frac{\etas^2a^dL_{\parallel}^{d-1}}{\kappa
L^{d-2}}\Upsilon_{\rm C}\left(\frac{L}{l_{\Omegas}}\right),\label{FCbis}\\
\Upsilon_{\rm C}\left(u\right )&=&\frac{1}{4u^2}\int_{\bf
q}\!\!q\left(\coth
q-1\right)\label{upsilonc}\\&-&\frac{1}{4u^2}\int_{\bf
q}\!\!\sqrt{q^2+u^2}\left[\coth
\left(\sqrt{q^2+u^2}\right)-1\right ].\nonumber \end{eqnarray}
The scaling function $\Upsilon_{\rm C}$ is plotted
in Fig.~\ref{upsiloncplot}. To achieve the exact computation of
$F_{\rm C}$, the integral over the in-plates $\bf q$ modes in
 (\ref{upsilonc}) is performed. After some algebra we obtain: If
$d=1$, $\Upsilon_{\rm
C}=(1-u[\coth(u)-1])/u^2$, if $d=3$, $\Upsilon_{\rm C}=[
3\zeta(3)-\pi^2-6(u^2-u\log(1-e^{2u})+3{\rm
Li}_{2}(e^{2u})]/(48\pi u^2)$, where $\zeta$ is the Rieman Zeta
function and $\rm Li_{2}$ the dilogarithm
function : ${\rm Li}_2(x)=\sum_{k=1}^\infty x^k/k^2$. Finally for the $d=2$ case  $\Upsilon_{\rm
C}=2\sum_{n}[K_{0}(2n u)+K_{2}(2nu)]$, where $K_{n}$ is the
$n^{th}$ K-Bessel function~\cite{gradstein}.

Notice that this
last calculation can actually be bypassed, if one already knows
the thermal Casimir force the left plate would feel if the soft
medium had a finite correlation length $\xi$, precisely if the
elastic Hamiltonian were: ${\cal H}=(\kappa/2)\!\int
[\vec{\nabla} \phi(\vec{R})]^{2}+[\phi(\vec{R})/\xi]^2
d^d{\vec{R}}$. Denoting this force $\tilde F_{\rm
Casimir}(L/\xi)$, we can show that
Eqs.~(\ref{FCbis},\ref{upsilonc}) can be recast into~\cite{bafg}:
\begin{equation} F_{\rm
C}=\frac{\etas^2a^d\Omegas^{-1}}{2\gamma\kt}\left[{\tilde F}_{\rm
Casimir}\left (0\right)-{\tilde F}_{\rm
Casimir}\left(\frac{L}{l_{\Omegas}}\right)\right],
\label{FCbypass}
\end{equation}
where ${\tilde F}_{\rm
Casimir}\left (0\right)$, of course corresponds to the Casimir
force the plate experiences in a scale-free
fluctuating medium, c.f. Eq.~(\ref{casimirforce}).

{\it Qualitative approach}:
Eq.~(\ref{FCsumFA}) shows clearly that  $F_{\rm C}$ is the
result of the  incoherent sum over frequencies and positions
of the forces produced by localized  shakers of the
--A-- type. Note that the sum over the shaking frequencies is
weighted by the noise power spectrum
$S_{\omega}$.
This decomposition is now used to justify first the sign
of the mean force, then the algebraic decays of the mean
force as observed in Eq.(\ref{FCbis}) and Fig.~(\ref{upsiloncplot}).

Only the elementary monochromatic shakers
localized at  distances $|x|<L$ on the left hand side of
the plate have a challenger in the $0<x<L$ inner region.
Referring to the discussion in Section IV-A, we know
that $F_{\rm A}^<>0$ and that $\left |F_{\rm A}^>(x,\omega)\right
|<\left |F_{\rm A}^<(-x,\omega)\right|$. Moreover, $S_{\omega}$
is necessarily a positive quantity for all frequencies. It then
turns out that the net mean force is positive, for any
choice of $S_{\omega}$: the two objects attract each
other. Besides, $F_{\rm
A}^{>}(x,\omega)\sim -F_{\rm A}^<(-x,\omega)$ for shaker
located at $x<L/2$. This important property leads to
the cancellation of the short distance divergences in the first
term of equation (\ref{FCsumFA}). Finally,
$\left|F_{A}^>(x,\omega)\right|\ll
\left|F_{A}^<(-x,\omega)\right|$ if $|x|>L/2$, so that Eq. (\ref{FCsumFA})
can be approximated by $F_{\rm C}\sim2L_{\parallel}^{d-1}a^{-d}
\int_{\omega}S_{\omega } \int _ { L / 2 } ^ { \infty } F _ { A }^ <
(-x,\omega)\,dx $.

To sum up, the elastic field deformations produced by the
--A-- type shakers located at
$x<-L/2$ dominate the mean net force $F_{\rm C}$.

Without refering to any specific choice for $S_\omega$, 
the noise power spectrum is here characterized only by its
width $\Omega_{\rm s}$, see e.g. Eq. (13). Thus,  going to the $(x,
l_{\omega})$ variables and defining $L^{*}\equiv{\rm
max}(L,l_{\Omegas})$ we easily deduce:
\begin{equation}
F_{\rm
C}\sim\frac{\etas^2 a^d\Omegas^{-1}L_{\parallel}^{d-1}}{\gamma }
\int_{L^*}^{\infty}\frac{d\,l_{\omega}}{l_\omega^3}\!\!\int_{L}^{l_{\omega }
}\!\!\frac{dx}{x^{d-1}},
\label{FCapprox}
\end{equation}
since $F_{\rm A}^<(x,\omega)$ is
exponentially weak if $x>l_{\omega}$ and scales as $1/x^{d-1}$ if
$x<l_{\omega}$, c.f. Fig.~(9).

We can now integrate (\ref{FCapprox}) and distinguish the two
cases:

(i) $L/l_{\Omegas}\gg1$.
In this wide noise spectrum limit $F_{\rm
C}\sim(\etas^2 a^d\Omegas^{-1}/\gamma)L_{\parallel}^{d-1}/L^d$,
the thermal Casimir force scaling is obtained as anticipated in
section III.

(ii) $L/l_{\Omegas}\ll1$. In this case the noise correlation time
$\sim \Omegas^{-1}$ is larger than the field relaxation over
distances smaller than L. The number of elementary shakers
contributing to the force is substantially increased compared to
(i). Depending on the space dimension $d$, we obtained the three
scaling forms: if $d=3$, $F_{\rm
C}\sim(\etas^2a^3\kappa^{-1}) L_{\parallel}^2/L$, if
$d=2$, $F_{\rm
C}\sim(\etas^2a^2\kappa^{-1}) L_{\parallel}\log(l_{\Omegas}/L)$,
an finally if $d=1$, $F_{\rm
C}\sim(\etas^2a\kappa^{-1})l_{\Omegas}$.
 In agreement with the above exact calculations and with
 Fig.~\ref{upsiloncplot}.

{\it Fluctuations of the shaking-induced force:}
To conclude this section, we address briefly the question of the
fluctuations of the  force. In order to asses their
relative importance with respect to
the mean force value, we introduce the dimensionless ratio:
$F_{\rm C}/\Delta F$, where the mean squared deviation $\Delta
F^2$ is defined by: $\Delta F^2\equiv\langle F(t)^2\rangle-F_{\rm
C}^2$. Whereas $F_{\rm C}$ is the result of the competition
between the mean forces $F_{\rm C}^>$ and $F_{\rm C}^<$ of
opposite signs, the forces' fluctuations on the two sides add up
and $\Delta F^2=(\Delta F^{>})^2+(\Delta F^{<})^2$ with obvious
notations. Note that the
noise's four points correlation function is a priori required to
compute $\Delta F$. We make here one
more assumption and consider that the $\eta(\vec{R},t)$ are
Gaussian random variables with correlations given by
Eq.~(\ref{etac},\ref{powerspec}). This allows to fully
characterize the shaking process and consequently the variance of
the shaking-induced force. For sake of clarity we present without
additional details the scaling form of the mean over variance
ratio (ignoring logarithmic corrections):
\begin{equation} \frac{F_{\rm C}}{\Delta F}\sim
\left(\frac{L_{\parallel}}{L}\right)^\frac{d-1}{2}
\left\{
\begin{array}{ll}
\left(\frac{L}{l_{\Omegas}}\right )^{\frac{3-d}{2}} &{\rm if\,}
L/l_{\Omegas}\ll1\\
\left(\frac{l_{\Omegas}}{L}\right )^{\frac{d+1}{2}}&
{\rm if\,} L/l_{\Omegas}\gg1
\end{array}
\right .
\end{equation}
we have also assumed $d>1$, since the 1D case deserve a more
careful analysis as in the equilibrium context more carefully
studied in~\cite{bafg}. The above expressions shows that,
the present shaking induced force can not be a priori reduced
to its sole $F_{\rm C}$ value $F_{\rm C}$. Indeed, depending on
the geometrical aspect ratio $L_{\parallel}/L$ and on the noise
correlation time $\Omegas^{-1}$ the variance of the force can
dominate its mean value by orders of magnitude.
\subsection{Toward more complex dynamics} To what extent do the results
presented above apply to systems evolving according to more complex
dynamics? It is noticeable that only the explicit form of the
scaling functions $\Upsilon_{\rm X}$ requires a precise description of
the field dynamics. Conversely, all the qualitative analysis, as well as
the asymptotic expressions for the various forces, involve only the
definition of the dynamical length $l_{\omega}$. We thus expect that
they should be extendable to any other dynamic scheme relating
algebraically the spatial and the temporal relaxation scale.
\section{Summary and Outlook}
To summarize, a simple model has allowed us to show that the concept of
fluctuation-induced interactions can be extended  to complex fluids
driven to out of equilibrium.  Our main result concern identical
plate-like inclusions immersed in an homogenous  medium externally
driven by :--A-- a monochromatic localized shaking source located
between the plates, --B-- a monochromatic and homogenous shaking and
--C-- a noisy colored shaking. These external processes generate
effective forces on the plates without acting directly on them. The main
features of the mean forces induced by these three simple shaking are
summarized in the table below and compared to their equilibrium
thermal Casimir analog.
\begin{center}
\begin{tabular}{|c|c|c|c|}
Shaking&Amplitude  &Range  & Sign  \\ \cline{1-4}
{\rm A} &$\etas^2$&$l_{\omegas}$&repulsion \\
{\rm B} &$\etas^2$ &$l_{\omegas}$ &tunable  \\
{\rm C} &$\etas^2$  &power law  &attraction  \\
{\rm Thermal} &$\kt$  &power law &attraction
\end{tabular}
\end{center}
Five points are worth a highlight:
\begin{description}
\item
(i) Whereas a shaking at a single
frequency leads necessarily to short range interactions, noisy
shakings generate effective forces that decays algebraically with
the interplate distance.
\item
(ii) If the external shaking occurs only between the two plates
the resulting interaction is always repulsive. Conversely, when
distributed on the whole sample the external shakers generate
forces on both sides of the plates. The resulting net force is
attractive for spatially uncorrelated shaking. On the
contrary, with coherent shaking, interference phenomena between
the elastic medium distortions can lead to attractive or
repulsive interactions.
\item
(iii) The soft medium-plates coupling has been here modeled by Dirichlet boundary conditions. 
As in the equilibrium case
our results are expected to hold for other static boundary conditions representing strong interactions.
In contrast the description of situations of weaker coupling and/or cases where dynamic boundary conditions are required could yield to new phenomenology.
\item
(iv) The generalization of our results to an elastic medium characterized by an intrinsic relaxation scale $\xi$ (see the discussion above Eq. (34)) can be done without much effort following blindly the procedures described in this paper. The only qualitative difference in the results summarized in the above table is the forces' range: For shakings of the --A-- and --B-- type the range of the force becomes: $\min(\xi,l_\omega)$. In the noisy shaking case --C-- the force decay is not algebraic anymore but exponential for interplate distances larger than $\xi$.  
\item
(v) Beyond the derivation of their average values, we have also
shown that the shaking-induced forces are in general expected to
be strongly fluctuating quantities, whatever their precise
origin.
\end{description}
To conclude, let us recall that our description is clearly
extremely simplified and of course is not supposed to model
any  real system. Even within this
simple framework the extension of our result to
many moving objects remains complex. But, the analysis of
simplified situations such as the one described here could
certainly help understanding ordering and dynamic behavior of
inclusions in out-of equilibrium complex fluids.


\begin{references}
\bibitem{phase}
The phase shift with respect to the original
shaking process has been ignored for sake of clarity.
\end{references}
\end{document}